# Search for Magnetic Order in Superconducting $RuSr_2Eu_{1.2}Ce_{0.8}Cu_2O_{10}$

**J.W. Lynn**[1], **Y. Chen**[1,2], **Q. Huang**[1], **S. K. Goh**[3], **G. V. M. Williams**[3], and **J. L. Tallon**[3]

[1]NIST Center for Neutron Research,

National Institute of Standards and Technology, Gaithersburg, MD 20899

[2]Department of Materials Science and Engineering, University of Maryland, College Park,

MD 20742

[3]MacDiarmid Institute for Advanced Materials and Nanotechnology, Industrial Research,

P.O. Box 31310, Lower Hutt, New Zealand

Abstract

Neutron diffraction, polarized neutron transmission, and small angle neutron scattering have been used to investigate the crystal structure and nature of the magnetic order in a polycrystalline sample of $RuSr_2Eu_{1.2}Ce_{0.8}Cu_2O_{10}$. The sample was made with the Eu-153 (98.8%) isotope to reduce the high neutron absorption for the naturally occurring element. Full refinements of the crystal structure, space group *I4/mmm*, are reported. At low temperatures only a single magnetic peak is clearly observed in a relatively wide angular range. A sharp spin reorientation transition (SRT) is observed around 35 K, close to the superconducting transition temperature ($T_c$~40 K). Between the spin reorientation temperature and the Neel temperature of 59 K, additional magnetic reflections are observed. However, none of these can be simply indexed on the chemical unit cell, either as commensurate peaks or simple incommensurate magnetism, and the paucity of reflections at low T compels the conclusion that these magnetic Bragg peaks arise from an impurity phase. X-ray and neutron diffraction on the pressed pellet both show that the sample does not appear to contain substantial impurity phases, but it turns out that the magnetic impurity peaks exhibit strong preferred orientation with respect to the pellet orientation, while the primary phase does not. We have been unable to observe any magnetic order that can be identified with the ruthenate-cuprate system.

## I. Introduction

The properties of the ruthenate-cuprate class of materials have been of particular interest for a decade because of the high magnetic ordering temperatures for the Ru spins, which occur well above the superconducting temperature regime [1-16]. The $RuSr_2GdCu_2O_8$ material (Ru1212) has been investigated in the most detail, where the Ru spins are found to order antiferromagnetically at 136 K, well above the onset of superconductivity at 40 K, while the Gd spins order antiferromagnetically at 2.5 K [4] in a manner similar to the lanthanide magnetic order in a wide variety of cuprate superconductors [17]. The properties of the $RuSr_2R_2Cu_2O_{10}$ ($R$=rare earth) system of direct interest here (Ru1222) have been more elusive. This system has been reported to have a larger ferromagnetic component than the Ru1212 system, and multiple magnetic transitions in the range from 225 K to 20 K that have been difficult to interpret and understand [1-16, 18-25].

Magnetic neutron diffraction experiments have not been reported for the Ru1222 system because of the very high neutron absorption of Eu. We therefore have undertaken neutron diffraction measurements utilizing a lower absorbing isotope of Eu, in an effort to determine the nature of the magnetic order. Initially we obtained inconsistent and confusing results, whereby we carried out a series of additional diffraction experiments as a function of temperature and magnetic field, as well as polarized neutron transmission and small angle neutron scattering measurements, in an effort to determine the origin of these problems. We discovered that the sample contained sizable *single crystal* impurities, that ordered magnetically ~59 K, and these crystals exhibited a particular orientation with respect to the axis of the cylindrical pellet. At a lower temperature of 35 K a sharp spin reorientation transition is observed, coincidently in the vicinity of the superconducting phase transition. The majority $RuSr_2Eu_{1.2}Ce_{0.8}Cu_2O_{10}$ phase did not exhibit any preferred orientation in the powder, and that combined with the paucity of magnetic reflections at low temperature establishes that the observed magnetic ordering is not associated with this phase. Indeed we were unable to detect any magnetic ordering of the Ru moments in $RuSr_2Eu_{1.2}Ce_{0.8}Cu_2O_{10}$.

## II. Experimental Details

A polycrystalline sample was prepared in the usual way [8] in a pressed pellet form using the $Eu^{153}$ isotope to reduce the large neutron absorption from naturally occurring Eu. The pellet was approximately 1 cm in diameter and weighed ~1 gram. The neutron absorption for $Eu^{153}$ is still quite significant, and the sample was left in pellet form for most of the measurements to reduce absorption effects as well to prevent possible rotation of the particles when a magnetic field was applied.

Coarse resolution/high intensity diffraction experiments were carried out on the BT-2 and BT-7 triple axis neutron spectrometers, using pyrolytic graphite monochromator, filter, and analyzer (when employed). The neutron energy was chosen to be 14.7 meV. For the zero-field measurements the sample was sealed in helium exchange gas and placed in a closed cycle refrigerator with a low temperature capability of 4 K. Magnetic field measurements were carried out separately in a 7 T vertical field superconducting magnet. Polarized neutron transmission measurements were carried out on BT-2 using Heusler monochromator and analyzer, again at 14.7 meV (2.359 Å). A final polarized transmission measurement on the powdered sample was carried out on the NG-1 reflectometer that employs a wavelength of 4.75 Å and supermirrors for polarizers. Small angle neutron scattering measurements were collected on the NG-1 SANS instrument using a neutron wavelength of 6 Å, collecting data over a wave vector range of 0.004-0.04 Å$^{-1}$.

High resolution powder diffraction data were collected on the BT-1 spectrometer with monochromatic neutrons of wavelength 1.5403 Å produced by a Cu(311) monochromator and 2.0775 Å with a Ge(311) monochromator. Collimators with horizontal divergences of 15´, 20´, and 7´ arc were used before and after the monochromator, and after the sample, respectively. The intensities were measured in steps of 0.05° in the 2θ range 3°-168°. Data were collected for variety of temperatures from 298 K to 4 K to elucidate the magnetic and possible crystal structure transitions. The structural parameters were refined using the GSAS program [26], using neutron scattering amplitudes of 0.882, 0.484, 0.702, 0.721, 0.772, and 0.581 (×10$^{-12}$ cm) for $^{153}$Eu, Ce, Sr, Ru, Cu, and O, respectively. Initially the sample was kept in pellet form and continuously rotated to obtain the diffraction pattern, but these data were not of sufficient quality to fully refine. In the final set of diffraction measurements we crushed the pellet and enclosed the sample in a vanadium holder to obtain the data that were used in the final structural refinements.

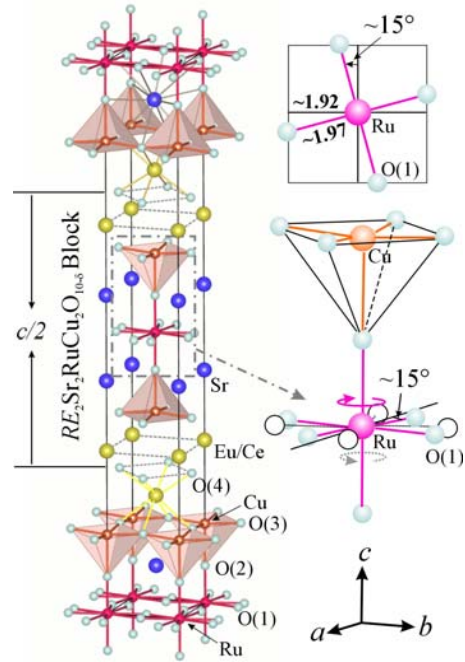

**Figure 1.** (color online) The structure for RuSr$_2$Eu$_{1.2}$Ce$_{0.8}$Cu$_2$O$_{10-\delta}$ (Ru1222). Left: structure built up by shifting the (½, ½, ½) RuSr$_2$Eu$_{1.2}$Ce$_{0.8}$Cu$_2$O$_{10-\delta}$ block from one another. Top-right: Rotating the RuO$_6$ octahedron ~15 degrees in the *ab*-plane modifies the four Ru-O in-plane distances from 1.92 to 1.97 Å to match the requirement of an average Ru-O distance ~1.96 Å for Ru$^{4+}$-O octahedron. Bottom-right: The disordered model obtained in the refinement suggests that there is an equal number of the clockwise (darker balls of O(1)) and counterclockwise (open circles) rotations.

## III. Results

### A. Crystal Structure

The structural refinements of RuSr$_2$Eu$_{1.2}$Ce$_{0.8}$Cu$_2$O$_{10-\delta}$ were carried out successfully on the powdered sample using the *I4/mmm* structural model, as shown in Fig. 1, which is isostructural with the compound $RuSr_2Gd_{1.3}Ce_{0.7}Cu_2O_{10-\delta}$ [10]. Some weak impurity peaks were found and could be indexed by lattice parameters close to the $SrRuO_3$ and $Sr_2EuRuO_6$ type compounds. These phases were, therefore, taken into account in the final calculations and their inclusion significantly improved the fit. Due to the small amount of the impurity phases—7.7% for the $SrRuO_3$-type (1:1:3) material and 5.6% for the $Sr_2EuRuO_6$-type double perovskite material [27]—only lattice parameters were allowed to be refined. In particular, we don't know the specific compositions for these phases and therefore can't relate their ordering temperatures or specific magnetic structures to our observations. The structural parameters and selected interatomic distances for a few representative temperatures are shown in Table 1. Figure 2 shows a plot of the diffraction data and fit at 4 K.

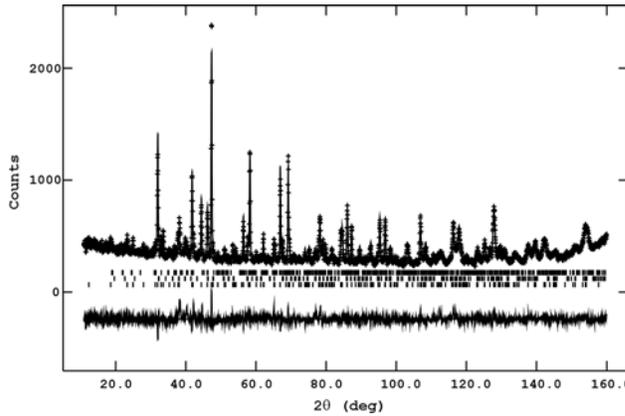

**Figure 2.** Observed (crosses) and calculated (solid curve) intensity profile at 4 K. The vertical lines on the bottom indicate the angular positions for the Bragg reflections for the $RuSr_2Eu_{1.2}Ce_{0.8}Cu_2O_{10-\delta}$ system, the middle for the 1:1:3 ($SrRuO_3$-like) impurity phase, and the top lines are for the $Sr_2EuRuO_6$-like double perovskite impurity phase. The lower part of the figure shows the difference plot, *I(obs)-I(calc)*.

No evidence of superstructures and/or orthorhombic distortions was observed. However, the large temperature factors in the *ab*-plane for the in-plane oxygen atoms O(1) of the $RuO_6$ octahedron suggest the presence of disordered rotations. A splitting of the O(1) site from 4*c* (1/2, 0, 0) to 8*j* (1/2, *y*, 0) in the *I4/mmm* symmetry was thus modeled. The structure was developed, as shown in Fig. 1, by shifting the (½, ½, ½) $RuSr_2Eu_{1.2}Ce_{0.8}Cu_2O_{10-\delta}$ blocks from one another. The $CuO_2$ layer is then above and below the $RuO_6$ octahedron layer. The Cu ions are coordinated by five oxygen atoms that form a pyramid with a longer apical Cu-O distance ~2.17 Å and four shorter in-plane Cu-O distances ~1.93 Å. The figure on the top-right in Fig. 1 shows the rotation of the $RuO_6$ octahedron by ~15 degrees in the *ab*-plane, which modifies the four Ru-O in-plane distances from 1.92 to 1.97 Å to match the requirement of an average Ru-O distance ~1.96 Å for the $Ru^{4+}$-O octahedron. The figure on the lower-right shows that the disordered model obtained in the refinement, and suggests that there are an equal number of clockwise (darker balls of O(1)) and counterclockwise (open circles) rotations. The refined structural parameters reported in the Table 1 are rotationally averaged results.

The refined occupancies for oxygen sites (O(1) and O(2)) surrounding the Ru are ~90%, *i.e.* the refined chemical formula is $RuSr_2Eu_{1.2}Ce_{0.8}Cu_2O_{9.6}$. Considering the ionic valences are 3+, 4+, 2+, and 2- for Eu, Ce, Sr, and O, respectively, the total valence for ionic Ru and 2Cu is 8.4+, in good agreement with the value of 8.45+ ($Ru^{3.91+}$ +$2Cu^{2.27+}$) obtained from the bond valence sum (BVS) [28] calculations given in Table 2.

In the corner-shared $RuO_6$ octahedra layer, the rotations of the $RuO_6$ octahedra are correlated, which induces superlattice parameters related to the tetragonal body-centered lattice by the transformation matrix (1, -1, 0/1, 1, 0/1, 1, 1) with $a_{sup}= \sqrt{2}\, a_{ave}$. The oxygen sites in the $RuO_2$ planes are then split into two positions with 50% occupancy for each. The presence of an oxygen vacancy can then break the correlation of the rotated $RuO_6$ octahedra. Figure 3 shows a proposed disordered rotational model in which the vacancies change the left-hand rotation (L) to be a right-hand rotation (R). With the present diffraction data and analysis, however, we are not able to identify a unique rotational model.

## B. Magnetic scattering

High intensity—coarse resolution measurements were carried out at a series of temperatures from 4 K to 200 K to search for the development of magnetic order. A portion of the diffraction pattern taken on BT-2 is shown in Fig. 4. The top portion of the plot shows the data obtained at 115 K, and the bottom portion of the data shows the subtraction of the 115 K data from the data collected at 5 K. In this subtraction process the paramagnetic background scattering evolves into Bragg peaks in the ordered phase, so in the subtraction there is a deficit of scattering away from the Bragg peak [29]. A single resolution-limited magnetic Bragg peak is observed at 42.5 degrees, indicating that long range magnetic order has developed in the sample. We remark that the scatter in the data around 50° and 60° is due to the subtraction of high intensity structural peaks which have a slight shift in their position and small change in the mean-square atomic displacements (Debye-Waller factor) with temperature, and does not originate from a magnetic order.

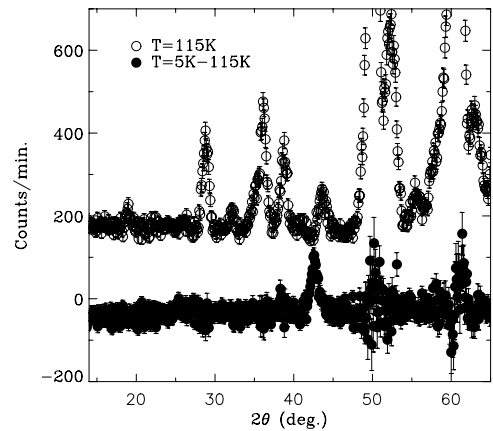

**Figure 4.** A portion of the high-intensity/coarse resolution data obtained at 115 K (top), and the magnetic diffraction pattern (bottom) obtained by subtracting these data from the data collected at 5 K. The strongest Bragg peaks are off scale so that the magnetic peak can be seen. In this subtraction process the diffuse paramagnetic scattering evolves into Bragg peaks in the ordered phase, so in the subtraction there is a deficit of scattering away from the Bragg peak [29]. A single resolution-limited magnetic Bragg peak is observed at 42.5 degrees, indicating that long range magnetic order has developed in the sample. There are small changes evident in the strongest structural Bragg peaks due to the Debye-Waller factor and thermal expansion so that these intensities do not exactly subtract, but this is not evidence of any type of magnetic order.

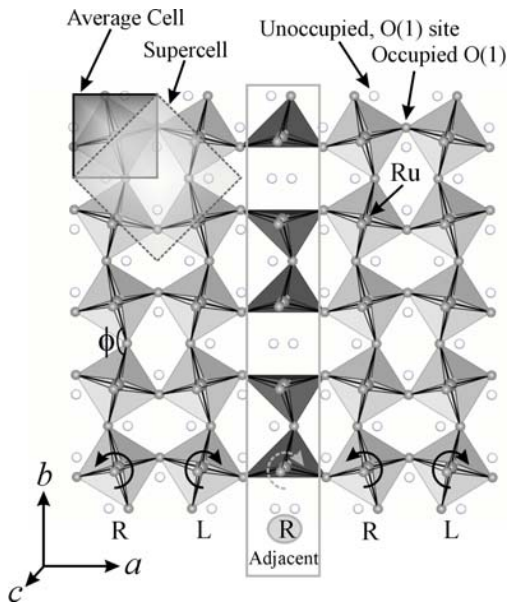

**Figure 3.** A proposed local-structure model of the disordered rotation for the $RuO_6$ octahedral layer in $RuSr_2Eu_{1.2}Ce_{0.8}Cu_2O_{10-\delta}$, showing the relationship between the average cell and the supercell. The oxygen vacancies can change the left-hand rotation (L) into a right-hand rotation (R). The Ru-O-Ru angle $\phi$ is ~150 degrees.

Figure 5 shows the magnetic scattering in detail. The low temperature magnetic peak is shown in Fig. 5a. At the intermediate temperature of 40 K (Fig. 5b) we see that the peak at 42.5° has decreased dramatically in intensity, while a new peak at 39.8° and a strong peak at 45.4° have appeared, indicating an abrupt change in the magnetic structure. At 60 K (Fig. 5c) just a very broad distribution of magnetic scattering is observed, indicating that we are above the magnetic ordering at this temperature. The temperature of the magnetic transition can be identified rather easily by observing the total intensity observed on the small angle neutron scattering (SANS) detector as a function of temperature, which is shown in Fig. 5d. The strong decrease in scattering around 60 K is due to critical magnetic scattering, which depletes the incident beam and thus reduces the intensity. We remark that a ferromagnetic transition would be expected to strongly increase the scattering in the small angle regime [30], so we identify this transition as antiferromagnetic in nature. No other magnetic transitions are observed in the SANS data over the temperature range explored.

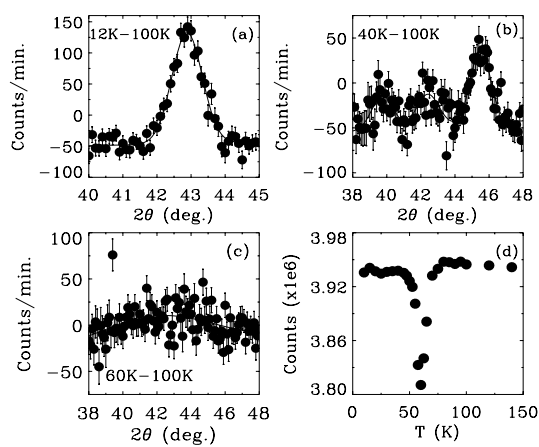

**Figure 5.** Magnetic scattering in detail. a) low temperature magnetic peak. b) at 40 K the peak at 42.5° has decreased dramatically in intensity, while a new peak at 39.8° and a strong peak at 45.4° have appeared, indicating an abrupt change in the magnetic structure. c) at 60 K only a broad distribution of magnetic scattering is observed, indicating that we are above the magnetic ordering at this temperature. d) total intensity observed on the small angle neutron scattering detector as a function of temperature. The sharp decrease (~3%) in scattering around 60 K is due to critical magnetic scattering, which depletes the incident beam and thus reduces the intensity. A ferromagnetic transition would be expected to strongly increase the scattering in the small angle regime, so we identify this transition as antiferromagnetic in nature.

The integrated intensities of the three magnetic peaks observed in Fig. 5 are shown in Fig. 6 as a function of temperature. The magnetic ordering temperature is determined to be 59 K, in excellent agreement with the SANS data. There is also a sharp (first order) spin reorientation transition around 35 K. We remark that no change in the positions of these three peaks is observed as a function of temperature.

It is instructive to compare the present results with those obtained on the related $RuSr_2Gd^{160}Cu_2O_8$ system [4], where there were two low angle magnetic peaks associated with the Ru magnetic order that were observed. These two peaks could be readily indexed on the chemical unit cell. The present single peak, on the other hand, presents a problem. With the long *c*-axis of the $RuSr_2Eu_{1.2}Ce_{0.8}Cu_2O_{10}$ structure, one would expect a series of magnetic peaks indexed as q=(0,0,2*l*), starting at an angle of ~9.5° and (approximate) multiples. Since none of these peaks is observed, the only way to explain the diffraction pattern based on the $RuSr_2Eu_{1.2}Ce_{0.8}Cu_2O_{10}$ crystal structure is to assume that the magnetic moments point along the *c*-axis. The magnetic intensities are proportional to

$$I \propto 1 - (\hat{q}\cdot\hat{M})^2 \quad ,$$

where $\hat{q}$ and $\hat{M}$ are unit vector in the direction of the reciprocal lattice vector and spin direction, respectively. With the moments parallel to the *c*-axis the intensities would be extinguished for these reflections. However, there would still be additional reflections that should be present but aren't observed. In addition, the single peak that is observed cannot be indexed in any simple way on the basis of the chemical unit cell. Thus a commensurate ferromagnetic or antiferromagnetic structure for $RuSr_2Eu_{1.2}Ce_{0.8}Cu_2O_{10}$ can be ruled out as giving rise to the observed magnetic peak. For an incommensurate magnetic structure, on the other hand, equally spaced satellite peaks about the structural peaks should be observed, so that we should see more magnetic peaks than in the case of a commensurate magnetic structure [31]. We can therefore also rule out an incommensurate magnetic structure as the origin of this lone magnetic Bragg reflection.

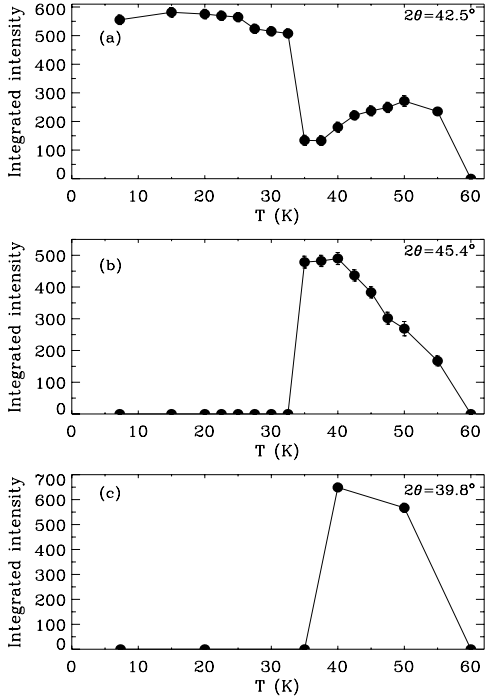

**Figure 6.** Integrated intensities of the three magnetic peaks observed in Fig. 5. The magnetic ordering temperature is determined to be 59 K. There is a sharp (first order) spin reorientation transition at 35 K. There is no change in the positions of these peaks as a function of temperature.

An additional difficulty that occurred in the diffraction measurements is that the intensity of the magnetic peak at low temperatures could be quite different from one thermal cycling to the next, or from one experiment to the next. Initially it was thought that the thermal path taken to low temperature was indeed an important factor, but numerous cycles did not reveal a pattern. The solution to these ambiguities is shown in Fig. 7. Figure 7a shows a rocking curve over a wide angular range for one of the strong structural reflections of the primary Ru1222 phase. For a properly randomized powder, of course, the observed intensity should be independent of the sample rotation, apart from absorption effects (neutron path length considerations) for the pellet itself. This is what is observed; the gradual intensity variation is due to the changing neutron path length in the pellet, with the minima occurring when the incident or scattered neutron path is parallel to the pellet. For the magnetic peak, on the other hand, we observe a very sharp peak, typical of a single crystal as shown in Fig. 7b. Here the "background" is ~35 counts/min, and then we see one sharp peak well above this. We remark that, in hindsight, we found these single crystal peaks have a preferred orientation with respect to the pellet, and typically in these triple-axis diffraction measurements the pellet was oriented to (approximately) bisect the incident and scattered beams. Hence it was serendipitous that we usually found a magnetic peak. Indeed, Fig. 7c shows that the intensity is also strongly dependent on the tilt; the resolution in the vertical direction is quite coarse and this width is comparable to the resolution. Then small shifts in the pellet with thermal cycling could cause the pellet to shift and thereby cause substantial changes in intensity. We remark that in our final experiment on the pellet, initially we detected no magnetic scattering, but then in searching for scattering we found a magnetic peak whose intensity, after properly orienting the sample, was as large as the strongest powder nuclear reflection in the Ru1222 phase. The clear and unambiguous conclusion is that the magnetic scattering originates from single crystals of an impurity phase embedded in the pellet. Since the Ru1222 phase is polycrystalline in nature, the magnetic ordering cannot be associated with this phase.

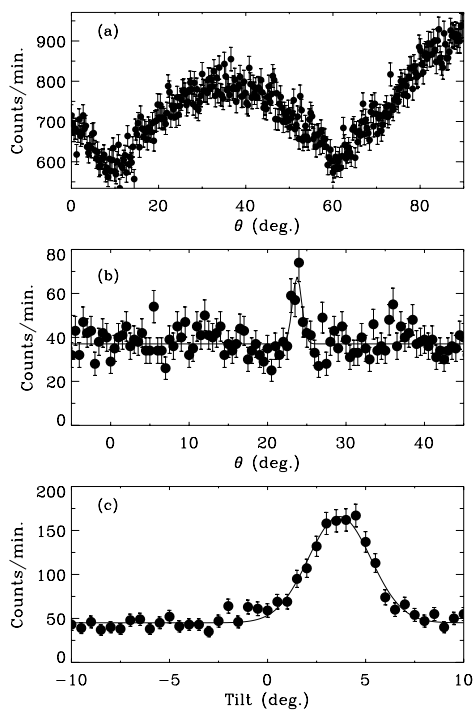

**Figure 7.** a) Rocking curve over a wide angular range for one of the strong reflections of the primary Ru1222 phase. The gradual intensity variation originates from neutron absorption in the pellet due to the changing neutron path length in the pellet, with the minima occurring when the incident or scattered neutron path is parallel to the pellet. b) Intensity of the magnetic peak as a function of sample rotation, which shows a very sharp peak, typical of a single crystal. We found that these single crystal peaks have a preferred orientation with respect to the pellet. c) Intensity of one of the magnetic impurity peaks as the sample is tilted. The single crystal nature of the magnetic impurity contrasts with the polycrystalline nature of the Ru1222 phase, demonstrating that the magnetic ordering cannot be associated with the primary phase.

Finally, we remark that we undertook two additional types of measurements. One was to measure the transmission of polarized neutrons through the sample, as we had done for the Ru1212 system [4], to search for a possible magnetic signal. The beam size was narrowed to a few mm in size to assure that all neutrons passed through the pellet, and to reduce the overall intensity on the detector. In our first measurement we found a small flipping ratio (R ~ 3) at low T, which increased up to the instrumental flipping ratio above ~60 K. This change in the flipping ratio suggested that there could be a ferromagnetic component associated with the magnetic ordering. No other change in transmitted polarization was detected up to room temperature. These data were collected before we had identified that the magnetic ordering originated from a single crystalline impurity phase. A later polarized neutron transmission measurement again found the instrumental flipping ratio (R ~ 26) at elevated temperatures, and a quite small change in the flipping ratio below ~60 K, which would suggest that this difference ("inconsistency") is again due to the serendipitous orientation of the magnetically ordered single crystals in the pellet. Finally, we note that we carried out a polarized transmission measurement on the NG-1 reflectometer on the sample after if had been crushed into a powder. The instrumental flipping ratio was ~50, and no change in the flipping ratio was observed from room temperature to ~6 K.

One other measurement we carried out was to apply a magnetic field to the sample in an attempt to detect an induced moment in the system. Data were collected for temperatures from 4 K to 100 K and fields up to 7 T. No induced magnetic moment was detected on the Ru1222 structural peaks, which puts a lower limit of ~0.25 $\mu_B$ on any induced Ru moment in the system.

## IV. Discussion

Our initial magnetic diffraction measurements on this sample quickly revealed a magnetic ordering in the system, and promised a swift determination of the basic magnetic structure. However, the inconsistent intensities on thermal cycling and the paucity of magnetic peaks suggested the possible formation of an exotic cooperative state, such as a spontaneous vortex lattice. Subsequent extensive neutron measurements did indeed reveal a rather unexpected, but not exotic, situation; a significant fraction of a magnetic impurity phase has formed in the system during preparation, and this phase exists as single crystals with a distinct preferred orientation with respect to the pellet. Full high resolution diffraction analysis of the crushed pellet indicates that ~13% of the sample consists of impurity phases. However, the small sample size and complexity of the sample and associated diffraction pattern does not allow a detailed determination of these impurity phases, and therefore we cannot relate the observed magnetic ordering to a particular impurity. It should be noted, however, that the polarized neutron transmission measurements did not observe any ferromagnetic ordering around the Curie temperature (160 K) of (pure) $SrRuO_3$, so it appears that the 1:1:3 impurity is not this composition.

There have been a variety of magnetic phases reported in the literature for this Ru1222 system. Our results make it clear that at least some of these are associated with impurities rather than the primary superconductor phase. Indeed, so far we have been unable to identify any long range magnetic order that is associated with the Ru1222 material. However, the Ru magnetic moment is small and difficult to detect, especially if the order is not fully long range in nature. Unfortunately, the nature of the magnetism in this ruthenate-cuprate system remains an unresolved issue, but we hope that the present results help in clarifying and correctly identifying the various magnetic transitions that have been reported in this system.

**Acknowledgments**

We would like to thank C. F. Majkrzak for his assistance on the NG-1 reflectometer. Funding in New Zealand supported by the New Zealand Marsden Fund and the Royal Society of New Zealand.

Table 1. Structural parameters for $Eu_{1.2}Ce_{0.8}Sr_2RuCu_2O_{10-\delta}$ (in the standard crystallographic notation) at 298 K and 4 K. Space group *I4/mmm*. Atomic positions are Eu/Ce 4*e* (0, 0, *z*), Sr 4*e* (0, 0, *z*), Ru 2*a* (0, 0, 0), Cu 4*e* (0, 0, *z*), O(1) 8*j* (½, *y*, 0), O(2) 4*e* (0, 0, *z*), O(3) 8*g* (0, ½, *z*), and O(4) 4*d* (0, ½, ¼).

| Atom | | 298 K | 4 K |
|---|---|---|---|
| | *a* (Å) | 3.8427(3) | 3.8326(2) |
| | *c* (Å) | 28.555(2) | 28.485(1) |
| Eu/Ce | z | 0.2953(3) | 0.2957(2) |
| | *B* (Å)$^2$ | 0.5(1) | 0.24(7) |
| Sr | z | 0.4227(3) | 0.4219(2) |
| | *B* (Å)$^2$ | 0.98(1) | 0.47(7) |
| Ru | *B* (Å)$^2$ | 0.4(2) | 0.3(1) |
| Cu | z | 0.1433(3) | 0.1431(2) |
| | *B* (Å)$^2$ | 0.4(1) | 0.16(6) |
| O(1) | y | 0.117(3) | 0.119(2) |
| | *B* (Å)$^2$ | 1.4(5) | 1.9(3) |
| | Occupancy | 0.45(2) | 0.46(1) |
| O(2) | z | 0.0672(3) | 0.0672(2) |
| | *B* (Å)$^2$ | 0.9(2) | 0.45(6) |
| | Occupancy | 0.92(3) | 0.86(1) |
| O(3) | z | 0.1495(2) | 1.4951(1) |
| | *B* (Å)$^2$ | 0.93(1) | 0.80(6) |
| O(4) | *B* (Å)$^2$ | 0.8(1) | 0.45(6) |
| | $R_p$ (%) | 4.34 | 4.70 |
| | $R_{wp}$ (%) | 6.39 | 6.37 |
| | $\chi^2$ | 1.090 | 1.767 |

Table 2. Calculated interatomic distances (Å) for $Eu_{1.2}Ce_{0.8}Sr_2RuCu_2O_{10-\delta}$ at 298 K and 4 K.

| | | 295 K | 4K |
|---|---|---|---|
| Cu-O(2) | ×~0.9 | 2.17(1) | 2.162(7) |
| Cu-O(3) | ×4 | 1.9294(6) | 1.9237(5) |
| **$V_{Cu}$ (e.u.)*** | | **2.27** | |
| Ru-O(1) | ×~3.6 | 1.973(3) | 1.970(2) |
| Ru-O(2) | ×~1.8 | 1.918(9) | 1.915(5) |
| **$V_{Ru}$ (e.u.)** | | **3.91** | |
| Sr-O(1) | ×~3.6 | 2.65(1)/3.24(1) | 2.660(6)/3.253(6) |
| Sr-O(2) | ×~3.6 | 2.733(1) | 2.7876(8) |
| Sr-O(3) | ×4 | 2.818(8) | 2.786(4) |
| Eu/Ce-O(3) | ×4 | 2.485(6) | 2.479(4) |
| Eu/Ce-O(4) | ×4 | 2.317(5) | 2.317(5) |

*N.E. Brese and M.O'Keeffe, Acta Cryst. B47, 192-197(1991).